\documentclass[a4paper,11pt]{article}
\usepackage{pos}

\title{Diffuse neutrino flux from coronal magnetospheric \\ current sheets of accreting black holes }
\ShortTitle{Diffuse neutrino flux from AGN coronae}

\author*[a]{D.~Karavola}
\author[a,b]{M~Petropoulou}
\author[c]{Damiano~F.~G.~Fiorillo}
\author[d]{L. Comisso,}
\author[d, e]{L. Sironi}

\affiliation[a]{National and Kapodistrian University of Athens, Panepistimiopolis, Athens, Greece}
\affiliation[b]{Institute of Accelerating Systems and Applications, Athens, Greece}
\affiliation[c]{Deutsches Elektronen-Synchrotron DESY,
Platanenallee 6, 15738 Zeuthen, Germany}
\affiliation[d]{Department of Astronomy and Columbia Astrophysics 
Laboratory, Columbia University, New York, NY 10027, USA}
\affiliation[e]{Center for Computational Astrophysics, Flatiron Institute, New York, NY, 10010, USA}

\emailAdd{dkaravola@phys.uoa.gr}
\emailAdd{mpetropo@phys.uoa.gr}

\abstract{Non-jetted AGN exhibit hard X-ray emission with a power law spectrum above $\sim$2~keV, which is thought to be produced through Comptonization of soft photons by electrons and positrons (pairs) in the vicinity of the black hole. The origin and composition of this plasma source, known as the corona, is a matter open for debate.
Our study focuses on the role of relativistic protons accelerated in black-hole magnetospheric current sheets in the neutrino production of AGN coronae. We present a model that has two free parameters, namely the proton plasma magnetization $\sigma_{\rm p}$, which controls the peak energy of the neutrino spectrum, and the Eddington ratio $\lambda_{\rm Edd}$ (defined as the ratio between X-ray luminosity $L_{\rm X}$ and Eddington luminosity $L_{\rm Edd}$), which controls the amount of energy transferred to secondary particles.
Furthermore, we combine our coronal model with an AGN population in order to provide a prediction for the diffuse neutrino flux measured on Earth. We compare our results with the observational data by IceCube and we find a satisfactory agreement on both the flux value and the slope of the neutrino distribution when we assume a $\sigma_{\rm p}$ value of $10^5$ for all the sources in our sample.}

\ConferenceLogo{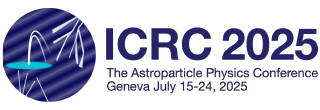}

\FullConference{39th International Cosmic Ray Conference (ICRC2025)\\
 14–24 July 2025\\
Geneva, Switzerland\\}
\begin{document}
\maketitle

\section{Introduction}

IceCube has detected a neutrino signal originating from the Seyfert II type galaxy with a significance of 4.2$\rm \sigma$ \cite{IceCube-NGC1068}. This detection has acted as a motivation for a lot of studies to revisit past radiation models for Active Galactic Nuclei (AGN) by adding a luminous proton component able to produce neutrinos. One such model is the \textit{coronal} one. By the term AGN corona, we usually refer to a compact region in the vicinity of the black hole that consists of energetic electrons, which upscatter UV photons emitted by the accretion disk to X-ray energies.  

However, neutrino observations point to the presence of protons in the coronal environment. Magnetic reconnection -- a process that dissipates magnetic energy and efficiently transfers it to non-thermal particles -- can accelerate a fraction of the aforementioned protons to highly relativistic energies \cite{chernoglazov_high-energy_2023, com2024}. Afterwards, protons may interact with coronal X-ray photons and produce neutrinos through photohadronic (p$\gamma$) interactions. Within such scenarios, questions about the energy and luminosity of the produced neutrinos will inevitably arise. In this work, we introduce a photohadronic coronal model and examine the neutrino distribution that escapes the latter environment. 

\section{Coronal model}

We introduce a model describing the coronal environment in which the X-ray emission is a result of magnetic reconnection at large-scale current sheets, with a typical size $R$ of a few gravitational radii $r_g=GM_{\rm bh}/c^2$ \cite{Ripperda_2020, el_mellah_reconnection-driven_2023, nathanail_magnetic_2022, sironi_relativistic_2015}, with $M_{\rm bh}$ being the mass of the central black hole. In the reconnection layer the non-thermal leptons receive a significant fraction of the Poynting luminosity. If the aforementioned particles transferred their energy to X-ray photons through some Comptonization process, the X-ray luminosity of the corona would be roughly equal to the Poynting luminosity \citep[e.g.][]{sironi_relativistic_2015, 2019ApJ...880...37P}. As a result, we express the broadband X-ray luminosity of the corona $L_{\rm \rm X}$\footnote{In this work, by $L_{\rm X}$ we will always refer to the bolometric X-ray luminosity of the corona in the $[0.1~\rm keV, 100 \, keV]$ band. The AGN catalogs provide the 2-10~keV X-ray luminosity. We extrapolate the latter to obtain $L_{\rm X}$ by assuming a power-law spectrum; see Eq. \ref{eq:PL_cor}.} as 
    \begin{equation}
        L_{\rm \rm X}= \eta_{\rm \rm X} S_{\rm \rm P} A = \eta_{\rm \rm X} \frac{c}{4 \pi} E_{\rm \rm rec} B A
        \label{eq:Lx}
    \end{equation} 
where $\eta_{\rm \rm X} \sim 0.5$ is the fraction of the magnetic energy dissipated to the X-ray photon field, $S_{\rm \rm P}$ is the Poynting energy flux, $A$ is the surface of the reconnection layer (taken to be $4 \pi R^2$), $B$ and $E_{\rm \rm rec}=\beta_{\rm \rm rec} B$ are respectively the magnetic field and electric field strengths in the upstream region of the layer, with $\beta_{\rm \rm rec} \sim 0.1$ being the reconnection speed in units of the speed of light $c$ \cite{chernoglazov_high-energy_2023, zhang_origin_2023, zhang_fast_2021}. We adopt an observationally motivated X-ray photon spectrum,
    \begin{equation}
    \begin{array}{cc}
         n_{\rm \rm X}(E)=n_{\rm \rm X,0}E^{-2}, & 0.1~{\rm keV} \le E  \le 100~{\rm keV}, \label{eq:PL_cor}
    \end{array} 
    \end{equation}
 where $n_{\rm \rm X}(E)$ is the differential number density of photons in the corona, and $n_{\rm X,0}$ is a normalization obtained from $L_X = 4 \pi R^2 c \int_{0.1~\rm keV}^{100~\rm keV} n_X(E)\,  E \,dE$.  

Driven by Particle-In-Cell (PIC) simulations of relativistic magnetic reconnection we expect that a fraction of the protons present will be injected into the acceleration process. Because protons are energized by the same process as electrons, we also assume that the power in relativistic protons is a fraction of the Poynting power entering the current sheet, so the bolometric energy injection rate into relativistic protons can be written as 
    \begin{gather}
        L_{\rm \rm p} = \eta_{\rm \rm p} S_{\rm \rm P} A = \frac{\eta_{\rm \rm p}}{\eta_{\rm \rm X}} L_{\rm \rm X}.
        \label{eq:Lp}
    \end{gather}
with $\eta_{\rm \rm p}\in [0.1, 0.5]$.

We consider protons to be accelerated in a prior phase in the upstream region, before being injected into the coronal environment, where they can cool due to p$\gamma$ interactions. As a result, we assume that the differential spectrum of protons injected into the corona per unit time by the acceleration process is also described by a broken power law \cite{chernoglazov_high-energy_2023, zhang_fast_2021, zhang_origin_2023},
\begin{equation}
       \frac{{\rm d^2}N_{\rm \rm p}}{{\rm d}\gamma_{\rm \rm p} {\rm d}t}= \dot{N}_{\rm \rm p,0} \left\{
        \begin{array}{cc}
           \gamma_{\rm \rm p, \rm br}^{-2} \gamma_{\rm \rm p}^{-1},  &  1\leq \gamma_{\rm \rm p} \leq \gamma_{\rm \rm p, \rm br} \\
            \gamma_{\rm \rm p}^{-3}, & \gamma_{\rm \rm p, \rm br} <  \gamma_{\rm \rm p} \leq \gamma_{\rm \rm \max}.
        \end{array}
        \right. \label{eq:dNdEdt}
    \end{equation}
where $\dot{N}_{\rm \rm p,0}$ is a normalization factor, $\gamma_{\rm \rm p, \rm br}$ is the break Lorentz factor, which is about equal to $\sigma_{\rm \rm p}$ \cite{com2024}. Here, the proton magnetization value is defined as 
\begin{equation}
\sigma_{\rm \rm p} = \frac{B^2}{4 \pi n_{\rm \rm p} m_{\rm \rm p} c^2},
\label{eq:sigma_p }
\end{equation}
with $n_{\rm \rm p}$ being the proton number density. The maximum energy of the proton distribution is set by the size of the corona (Hillas confinement criterion, $E_{\rm \rm p , \max} = e B R$ \cite{Hillas_1984}), or it is radiation-limited (i.e., the energy loss rate balances the acceleration rate).

In our previous study \cite{karavola25}, we showed that coronae with the same $\lambda_{\rm X, Edd}=L_{\rm X}/L_{\rm Edd}$ value behave in a similar way. More specifically, the luminosity transferred to neutrinos as a fraction of $L_{\rm X} $ is the same for coronae with the same $\sigma_{\rm p}$ and $\lambda_{\rm X, Edd}$ scaling linearly with the latter quantity until a saturation value. However, $\sigma_{\rm p}$ controls both the position and luminosity of the maximum of the neutrino distribution. As $\sigma_p$ increases both the aforementioned quantities increase, with the luminosity reaching a saturation limit eventually.

\section{Galaxy sample} \label{sec:galaxy_sample}
To test our model, we numerically calculated the neutrino spectra of six interesting sources, with NGC~1068 being one of them \cite{karavola25}. We found that the spectra predicted by our model consistently describe the IceCube detections or upper limits of 6 sources, with NGC~1068 being among them. Here, we expand our previous work by estimating the diffuse flux from a population of AGN coronae. We combine our coronal model with a mock catalog of AGN provided by \cite{georgakakis_forward_2020} so as to calculate the diffuse neutrino flux and compare it with IceCube observations.

The AGN mock catalog consists $\sim 10^5$ sources with $L_{\rm X}  \in [10^{42}, 10^{46}]~\rm erg/s$ in the 2-10keV energy band. Apart from the galaxy X-ray luminosity, the catalog also provides the stellar mass $M_{\rm stellar}$ and the redshift $z$ of each entry. We adopt the authors' relation for the central black hole mass which reads,
\begin{equation}
    M_{\rm bh}=2 \cdot 10^{-3} M_{\rm stellar}
\end{equation}

\begin{figure}
    \centering
    \includegraphics[width=0.9\linewidth]{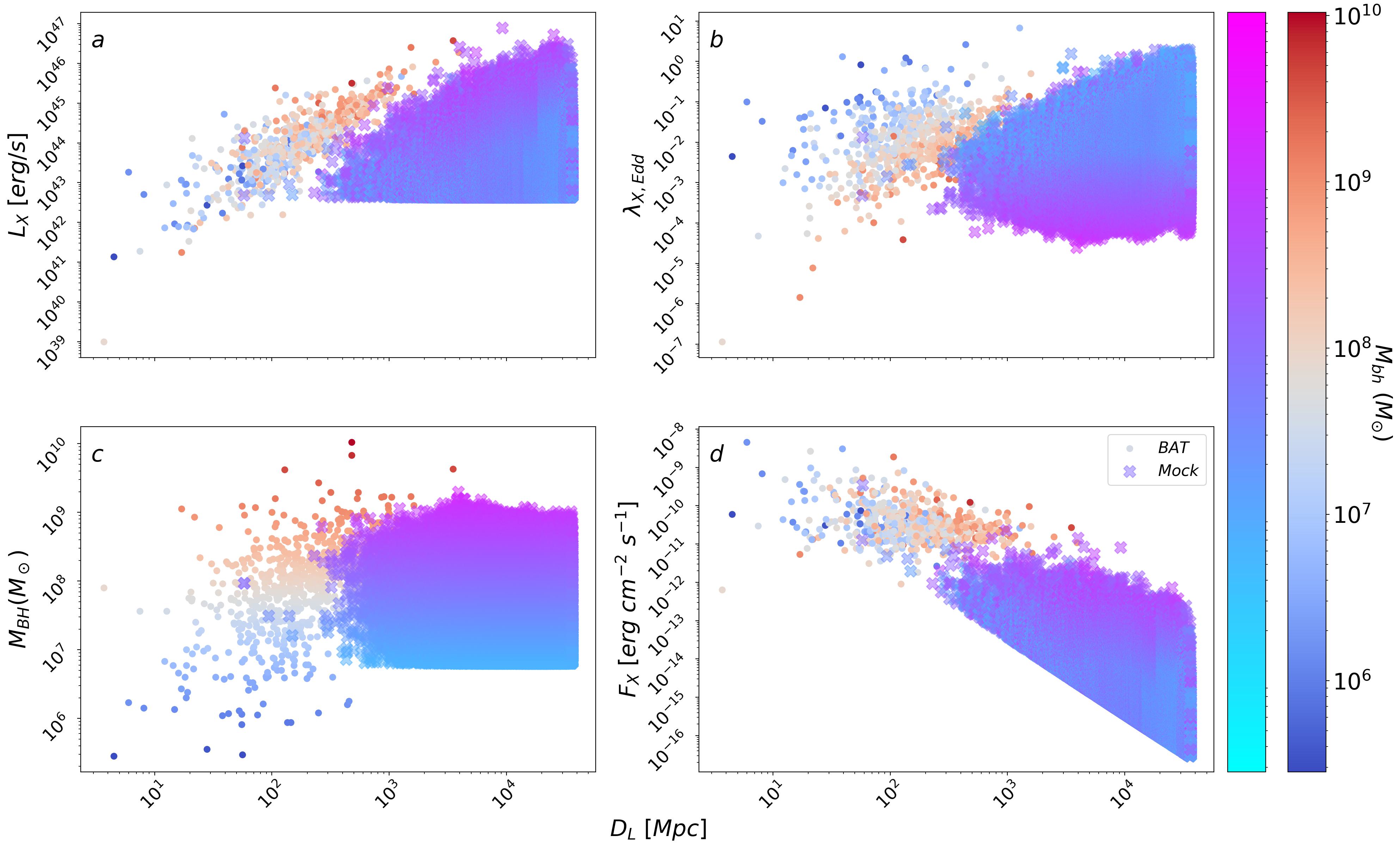}
    \caption{Scatter plots showing the X-ray luminosity distribution (panel a), the X-ray Eddington ratio (panel b), the black hole mass (panel  c), and the X-ray flux (panel  d) as a function of luminosity distance for sources from the BAT catalog (colored circles) and the mock AGN catalog (colored crosses). Color shows the black hole mass for both populations (see colorbars).}
    \label{fig:catalog_comparison}
\end{figure}

However, the simulated sample only includes sources with $z>0.5$. Therefore, to account for AGN in the local Universe, we supplement the mock catalog with the Swift-BAT 70-month catalog by \cite{ricci_bat_2017} which provides us with 677 non-blazar AGN (including NGC~1068). In Fig.~\ref{fig:catalog_comparison}, we display the parameter space for the quantities of importance to our model, such as the X-ray luminosity $L_{\rm X} $, the X-ray Eddington ratio $\lambda_{\rm X, Edd}$, the black hole mass $M_{\rm bh}$, and the X-ray flux $F_{\rm X}$ of the source as a function of its distance $D_L$ (in Mpc). One can notice that the luminosity, X-ray Eddington ratio and black hole mass distributions (panels a, b and c) of the two populations fall into the same range of values. However, galaxies located at greater distances (crosses) have lower fluxes (panel d) compared to the BAT sources (circles) as expected, since $F_{\rm X} \propto L_{\rm X}  D_L^{-2}$. One would expect that galaxies that are farther away would contribute less to the diffuse neutrino flux. However, we will show that this is not the case because the peak neutrino flux depends not only on the distance of the source but also on $\lambda_{\rm X, Edd}$ which regulates the all-flavor neutrino production efficiency and $\sigma_{\rm p} $ as shown by \cite{karavola25}:

\begin{gather}
f = \frac{E_{\nu} F_{\rm \rm \nu+\bar{\nu}}}{F_{\rm X}} \approx 0.125 \min  \left \{1, 1.2 \frac{\lambda_{\rm \rm X, Edd, -2}} {\tilde{R}}\frac{\min \left(5 \sigma_{\rm p, 5}, 46.5\right) \rm TeV}{5\, \rm TeV} \right\} \label{eq:Fnu}
\end{gather} 

where $F_{\rm X} = L_{\rm X}/(4 \pi D_L^2)$ and the notation $q_{\rm X}  = q/10^{\rm X}$ was introduced.

\section{Diffuse neutrino flux}

In this section we estimate the total neutrino flux expected from all the sources in the combined galaxy sample described in Sec.~\ref{sec:galaxy_sample} based on our coronal model. 

The shape of the neutrino spectrum produced by our model depends mainly on $\sigma_{\rm p} $ while its normalization depends on $\sigma_{\rm p}, \lambda_{\rm X, Edd}$, $L_X$ and $D_L$, as shown in Eq.~\ref{eq:Fnu} \cite{karavola25}.
We created neutrino spectral templates with the code {\tt ATHE$\nu$A} \cite{Dimitrakoudis, mastichiadis_spectral_2005} using NGC~1068 as a prototype 
($L_{\rm X} =6.3 \cdot 10^{43}$~erg/s and $M_{\rm bh}= 6.7 \cdot 10^6~M_{\rm \odot}$), and assuming that all coronae have the same magnetization. We performed calculations for $\sigma_{\rm p} = 10^3, 10^5$ and $10^7$.

\begin{figure}
    \centering
    \includegraphics[width=0.62\linewidth]{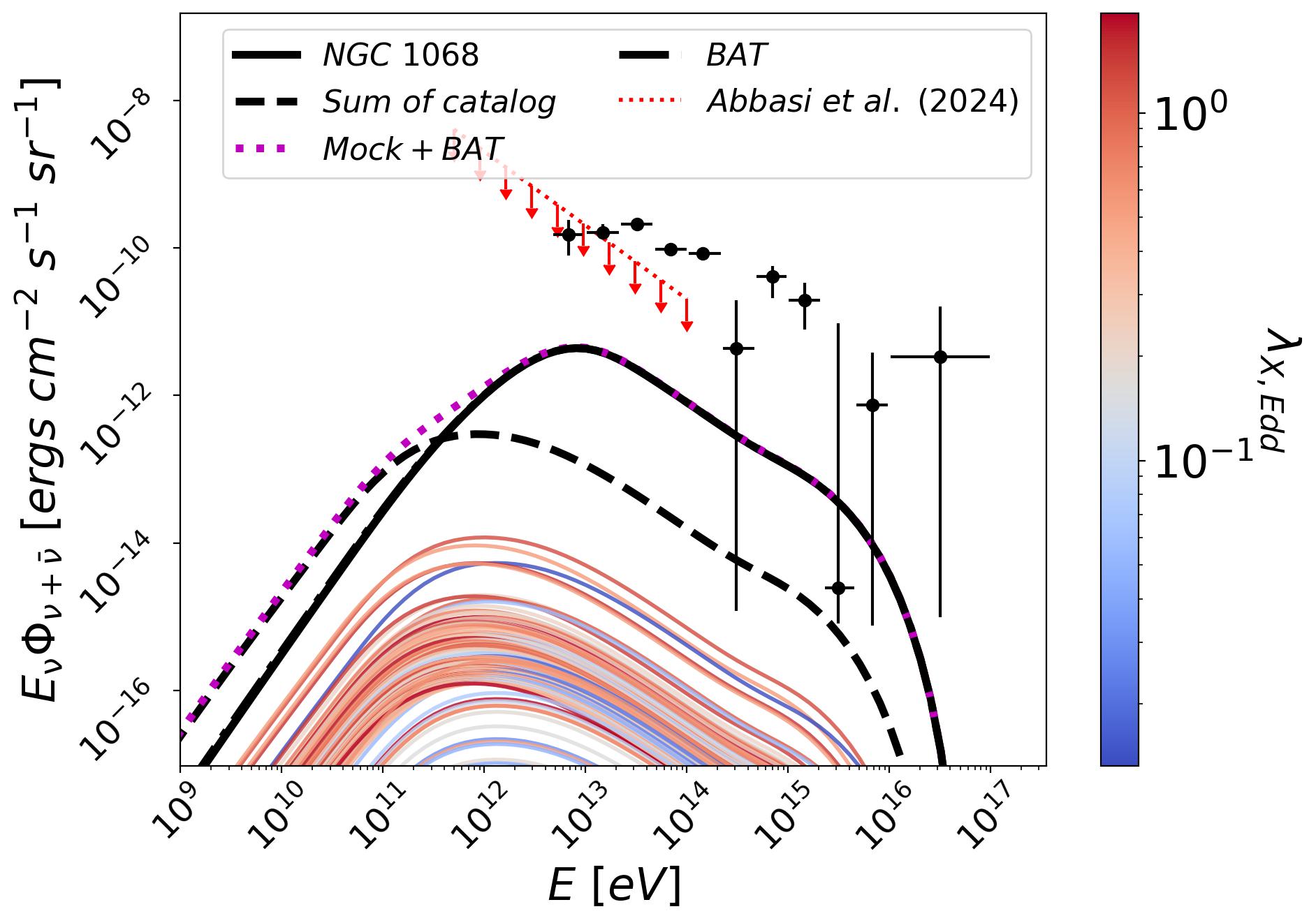}
    \includegraphics[width=0.62\linewidth]{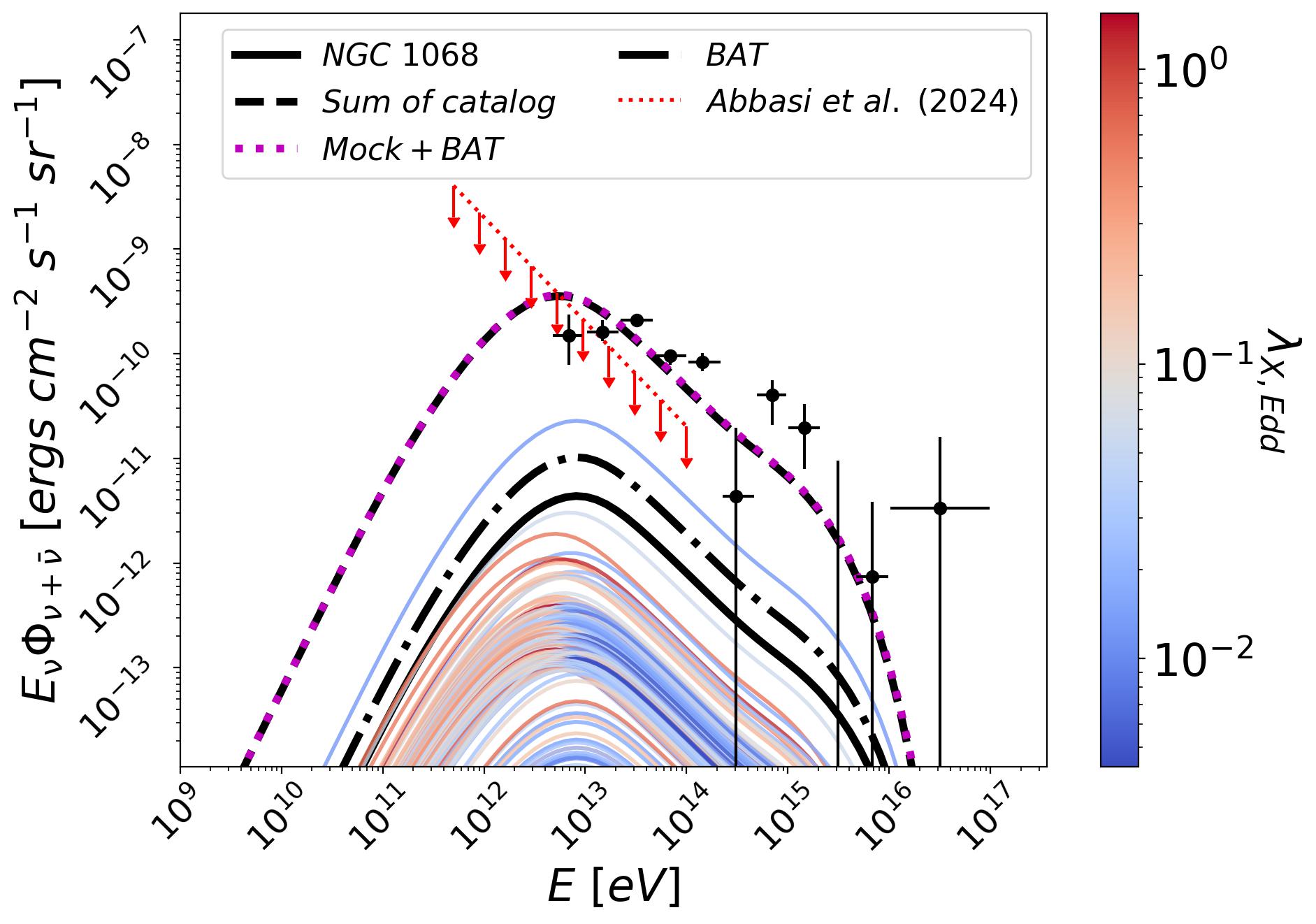}
    \includegraphics[width=0.62\linewidth]{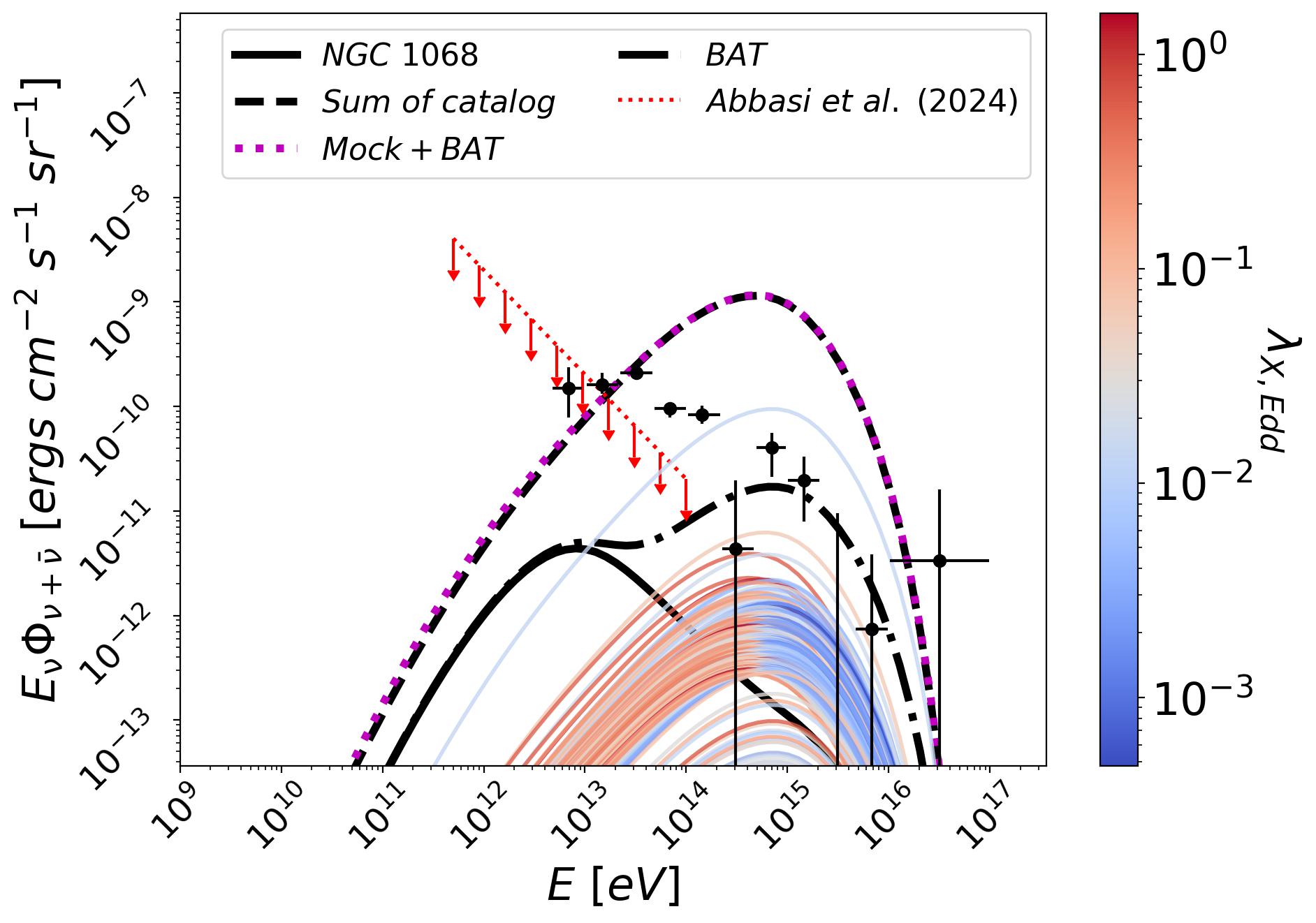}
    \caption{All-flavor neutrino energy flux of non-blazar AGN in the catalog of \cite{ricci_bat_2017} for three $\sigma_{\rm p} $ values, namely $10^3$ (top panel), $10^5$ (middle panel) and $10^7$ (bottom panel). Normalization scales as described by eq. \ref{eq:mock_sample}. Colored curves show neutrino spectra from a sub-sample of the mock catalog. Colors represent the ratio $\lambda_{\rm \rm X, Edd}$. The dashed black line shows the sum of all sources while the dash-doted black one represents the contribution of all the sources in the BAT catalog. Finally, the red dotted line with downward pointing arrows represents the neutrino flux upper limits for non-blazar sources from Table 1 of \cite{abbasi_search_2024} and the black markers indicate the diffuse neutrino flux observed by IceCube \cite{naab_measurement_2023}.}
    \label{fig:diff_spec}
\end{figure}

The neutrino flux of the $i$-th galaxy can be expressed in terms of the neutrino flux of NGC~1068 for a given $\sigma_{\rm p}$ value as
\begin{eqnarray}
(E_{\nu} F_{\nu+\bar{\nu}})_i  = \frac{1}{4\pi} \frac{(E_\nu L_{\nu+\bar{\nu}})_i}{4\pi D_L^2} = \frac{f_i L_{\rm X,i}}{(4\pi D_L)^2} = \frac{(E_\nu L_{\rm \nu+ \bar \nu})_{1068}}{(4\pi D_L)^2} \frac{f_i}{f_{1068}} \frac{L_{\rm X,i}}{L_{\rm X, 1068}}
\label{eq:mock_sample}
\end{eqnarray}
where $f$ is the neutrino-to-X-ray efficiency factor defined in Eq.~\ref{eq:Fnu}.

We calculate the diffuse neutrino spectrum by adding the contributions 
of individual sources from both catalogs. For each source we account for the redshift $z$ by shifting the neutrino energy as $E_{\rm \nu, obs}=E_{\nu}/(1+z)$. Moreover, the mock catalog of AGN accounts only for a fraction of 50 deg$^2$ of the sky, so we extrapolate it to half the sky to match the field-of-view of the BAT catalog,   
\begin{equation}
E_\nu^2 \Phi_{\nu + \bar{\nu}}(E_\nu)|_{\rm mock, obs}  = 50 \left(\frac{\pi}{180} \right)^2 \sum_{i} \left(E_{\nu+\bar{\nu}} F_{\nu+\bar{\nu}}\right)_i 
\end{equation} 

The results of the analysis described above are shown in Fig.~\ref{fig:diff_spec}, where the black solid line represents the spectrum of NGC~1068, as predicted by our model, and the colored solid lines show the fluxes of individual AGN from the BAT and mock catalogs, according to their X-ray Eddington ratio. Moreover, the black dashed line is the sum of all galaxies, including NGC~1068, while the red dotted line represents the neutrino flux upper limits for non-blazar sources given in Table 1 of Ref.~\cite{abbasi_search_2024}. 

As $\sigma_{\rm p} $ increases, the protons in the distribution that carry most of the total energy become more energetic. As a result,  the peak energy of the neutrino spectrum shifts to higher values. This was also demonstrated for individual sources in \cite{karavola25}. 
Additionally, the maximum neutrino energy flux increases with $\sigma_{\rm p}$ because the efficiency of neutrino production scales linearly with $\sigma_{\rm p} $ until it saturates (proton calorimetric limit). As a result, as long as protons in a corona are not calorimetric and the proton magnetization is higher, we expect more power to be transferred into neutrinos. Furthermore, by taking a closer look at Fig.~\ref{fig:diff_spec}, one can compare the results of the analysis performed in this work with the diffuse neutrino emission observed by IceCube, plotted with black markers. We find that $\sigma_{\rm p} =10^3$ underestimates the observed neutrino flux (top panel), while $\sigma_{\rm p} =10^7$ exceeds the observed spectrum (bottom panel). However, the case of $\sigma_{\rm p} =10^5$ (middle panel) seems to describe the observational data satisfactorily well, as it captures both the flux and the slope of the power-law tail. Moreover, we point out that in our analysis NGC~1068 remains among the brightest contributors to the stacked neutrino flux from the BAT catalog. 

\section{Discussion}

In this work, we combined our coronal model, previously introduced in a separate study, with two AGN catalogs, an observational one, namely the 70-month BAT catalog, and a mock AGN catalog introduced by \cite{georgakakis_forward_2020}. Our aim was to calculate the diffuse neutrino flux as predicted by the model and compare it to the IceCube 10-year diffuse neutrino flux \cite{naab_measurement_2023}.

The model depends on two free parameters. The first one is the X-ray Eddington ratio $\lambda_{\rm X, Edd} \propto L_{\rm X}/M_{\rm bh}$ which can be estimated for each source since both the X-ray luminosity $L_{\rm X}$ and the black hole mass $M_{\rm bh}$ can usually be retrieved from the literature. The second free parameter is the proton magnetization of the corona, $\sigma_{\rm p}$, which essentially determines the peak energy of the neutrino spectrum. 

In order to calculate the diffuse neutrino emission of the combined AGN catalog, we summed the contributions of  all individual sources. To calculate the latter, we used templates for the neutrino distribution shape, which only depends on the $\sigma_{\rm p}$ value. We renormalized the templates based on the X-ray luminosity and the neutrino production efficiency of each source contrary to previous studies in which only the $L_{\rm X}$ scaling  was employed \cite{padovani_neutrino_2024}. We find that if AGN coronae have $\sigma_{\rm p}=10^5$ the predicted diffuse neutrino spectrum describes the IceCube data satisfactorily in terms of flux and spectral shape. It is also worth mentioning that in the coronal environment, the magnetic field can reach values as high as a few $\sim$kG. In such strong fields, mesons produced by p$\gamma$ interactions can cool efficiently before decaying. As a result, the high-energy tail of the neutrino spectrum is expected to appear steeper, as shown in \cite{karavola25}. However, this is an effect we did not account for in this study, but will be the scope of our future work. 

\bibliographystyle{JHEP}
\bibliography{ICRC2025_template/bibliography} 

\end{document}